\documentclass[12pt,prd,tightenlines,nofootinbib,showpacs,showkeys]{revtex4}
\newcommand{\be}{\begin{equation}}
\newcommand{\ee}{\end{equation}}
\usepackage{bm}
\usepackage{graphics}
\usepackage{rotating}
\usepackage{epsfig}
\begin{document}
\title{\begin{flushright}{\rm\normalsize HU-EP-08/03}\end{flushright}
Relativistic description
of the double charmonium\\
production in $e^+e^-$ annihilation}
\author{D.Ebert}
\affiliation{Institut f\"ur Physik, Humboldt--Universit\"at zu Berlin,
Newtonstr. 15, D-12489  Berlin, Germany}
\author{R. N. Faustov}
\author{V. O. Galkin}
\affiliation{Russian Academy of Sciences, Dorodnicyn Computing
Centre, Vavilov Street 40, Moscow 119333, Russia}
\author{A. P.
Martynenko}
\affiliation{Institut f\"ur Physik, Humboldt--Universit\"at zu Berlin,
Newtonstr. 15, D-12489  Berlin, Germany}
\affiliation{Samara State University, Pavlov Street 1, Samara 443011,
Russia}

\begin{abstract}
New evaluation of the relativistic effects in the double production
of $S$-wave charmonium states is performed on the basis of perturbative
QCD and the relativistic quark model. The main improvement consists in
the exact account of properties of the relativistic meson wave
functions.
For the gluon and quark propagators entering the production vertex
function we use a truncated expansion
in the ratio of the relative quark momenta to the center-of-mass
energy $\sqrt{s}$ up to the second order.
The exact relativistic treatment of the wave functions makes all
such  second order terms convergent, thus
allowing the reliable calculation of their contributions to the
production cross section. Compared to the nonrelativistic calculation
we obtain a significant increase of the cross sections for
the $S$-wave double charmonium production. This brings  new
theoretical results in good
agreement with the available experimental data.
\end{abstract}

\pacs{13.66.Bc, 12.39.Ki, 12.38.Bx}

\keywords{Hadron production in $e^+e^-$ interactions, Relativistic quark model}

\maketitle

The production processes of mesons and baryons containing heavy $b$
and $c$ quarks in different reactions are under intensive study at
present \cite{BJ,BFY,UFN1,QWG}. The experimental investigation of
the double charmonium production in $e^+e^-$ annihilation by BaBar
and Belle Collaborations revealed a discrepancy between the measured
cross sections and theoretical results obtained in the
nonrelativistic approximation in QCD \cite{Belle,BaBar,BL1}. Various
efforts have been undertaken to improve the theoretical
calculations. They include the evaluation of radiative corrections
of order $\alpha_s$ and the investigation of relativistic effects
connected with the relative motion of the heavy quarks forming the
vector and pseudoscalar quarkonia
\cite{BL1,Chao,Chao1,Ma,BC,BLL,ZGC,Bodwin2,EM2006,Ji,He,AVB,Bodwin4}.
As a result, the difference between theory and experiment for the
value of the center-of-mass energy $\sqrt{s}=10.6$ GeV was
essentially decreased \cite{BL1,BLL,ZGC,EM2006}. Moreover, the new
theoretical analysis carried out in Refs.\cite{Bodwin4,Bodwin3}
shows that the inclusion of order $\alpha_s$ and relativistic
corrections decreases the discrepancy between theory and experiment
at the present level of precision. But despite this fact there
exists the frequently debated question connected with the
calculation of the relativistic corrections in the production cross
section. It is related to the determination of the specific
parameter $\left\langle{\bf p}^2\right\rangle$=$\int{\bf
p}^2\Psi_0^{{\cal P}, {\cal V}}({\bf p})d{\bf p}/(2\pi)^3$ emerging
after the expansion of all quantities in the production amplitude in
the relative quark momenta ${\bf p}$ and ${\bf q}$
\cite{apm2005,Bodwin1,EM2006,Bodwin3}, where $\Psi_0^{\cal V,P}$ are
the vector and pseudoscalar charmonium wave functions in the rest
frame. The divergence of this integral required the use of a
regularization procedure (dimensional regularization is commonly
used) which led to a definite uncertainty of the evaluation.
Moreover, the large value of the relativistic contribution obtained
in the previous studies \cite{BL1,EM2006} evidently rises a question
about the convergence of the expansion in the heavy quark velocity.
In this letter we propose an alternative approach to the calculation
of relativistic effects based on the relativistic quark model
\cite{savrin,rqm1,rqm2,rqm3,rqm4} and perturbative QCD. It uses a
truncated expansion in relative momenta ${\bf p}$ and ${\bf q}$ and
thus avoids divergent integrals in the relativistic contribution of
the second order.

\begin{figure}
\centering
\includegraphics[width=5.cm]{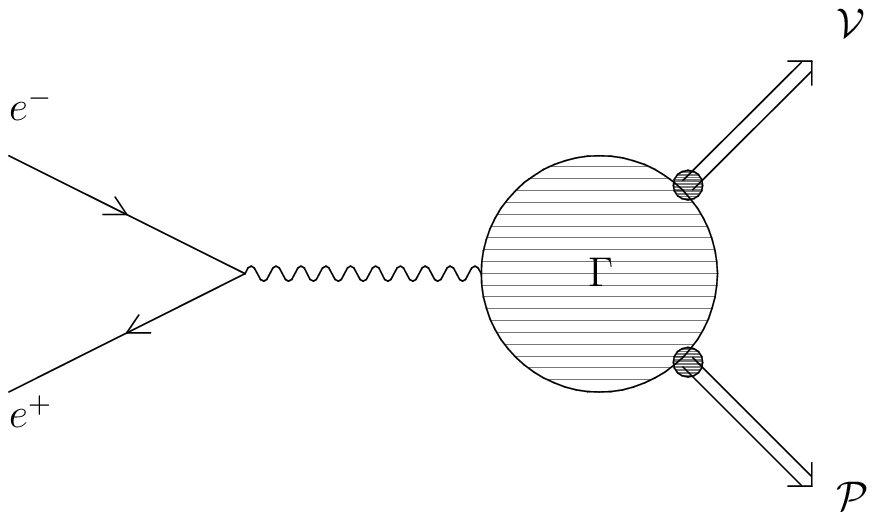}\hspace*{0.4cm}
\includegraphics[width=11cm]{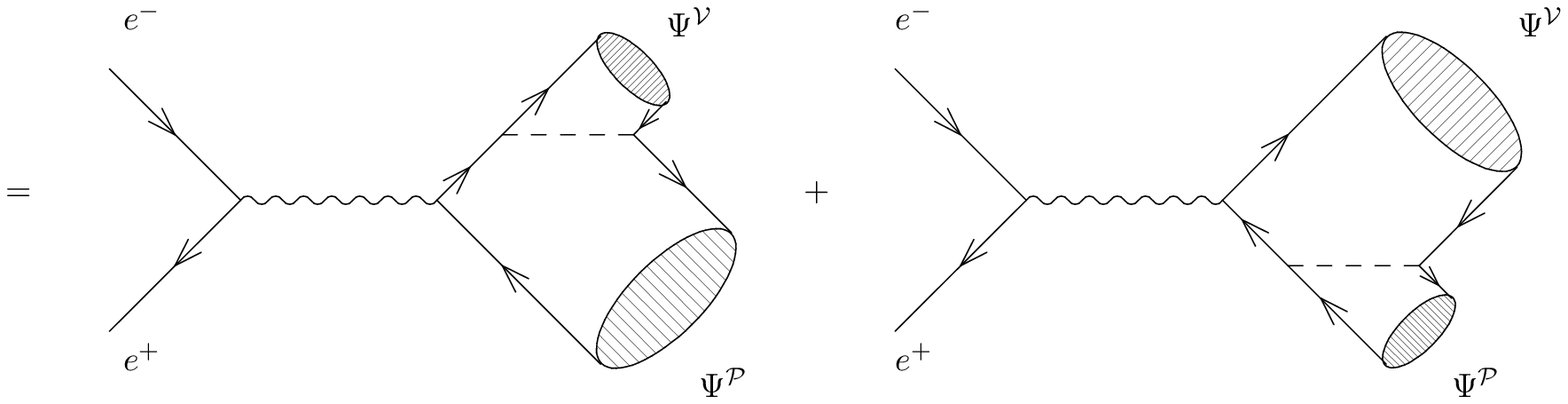}
\caption{The production amplitude of a pair of charmonium states
(${\cal V}$ denotes the vector meson and ${\cal P}$ the pseudoscalar meson)
in $e^+e^-$ annihilation.
The wave line shows the virtual photon and the dashed line corresponds
to the gluon. $\Gamma$ is the production vertex function.}
\end{figure}

Define the four momenta of the produced $c,\bar c$ quarks forming the
vector and pseudoscalar charmonia in terms of total momenta $P(Q)$ and
relative momenta $p(q)$ as follows:
\begin{equation}
p_{1,2}=\frac{1}{2}P\pm p,~~(p\cdot P)=0;~~q_{1,2}=\frac{1}{2}Q\pm q,~~(q\cdot Q)=0,
\end{equation}
where $p=L_P(0,{\bf p})$, $q=L_P(0,{\bf q})$ are the four-momenta
obtained from the rest frame four-momenta $(0,{\bf p})$ and $(0,{\bf
q})$ by the Lorentz transformation to the system moving with the
momenta $P,Q$. Then the production amplitude of the $S$-wave vector
and pseudoscalar charmonium states, shown in Fig.1, can be presented
in the form \cite{F1973,rqm5,EM2006}:
\begin{eqnarray} \label{M}
{\cal M}(p_-,p_+,P,Q)&=&\frac{8\pi^2\alpha\alpha_{s}{\cal
Q}_c}{3s}\bar v(p_+)\gamma^\beta u(p_-)\cr &&\times\int\frac{d{\bf
p}}{(2\pi)^3}\int\frac{d{\bf q}}{(2\pi)^3} Sp\left\{\Psi^{\cal
V}(p,P)\Gamma^\nu(p,q,P,Q)\Psi^{\cal P}(q,Q)\gamma_\nu \right\},
\end{eqnarray}
where $\alpha_{s}$ is the QCD coupling constant, $\alpha$ is the
fine structure constant, ${\cal Q}_c$ is the $c$ quark electric
charge. The relativistic wave functions of the bound quarks
$\Psi^{\cal V, \cal P}$ accounting for the transformation from the
rest frame to the moving one with four momenta $P,Q$ are
\begin{eqnarray}
\Psi^{\cal V}(p,P)&=&\frac{\Psi_0^{\cal V}({\bf p})}{\left[\frac{\epsilon(p)}{m}
\frac{(\epsilon(p)+m)}{2m}\right]}\left[\frac{\hat v_1-1}{2}+\hat v_1\frac{{\bf
p}^2}{2m(\epsilon(p)+ m)}-\frac{\hat{p}}{2m}\right]\cr
&&\times\hat{\tilde\epsilon}^\ast(1+\hat v_1)
\left[\frac{\hat v_1+1}{2}+\hat v_1\frac{{\bf p}^2}{2m(\epsilon(p)+
m)}+\frac{\hat{p}}{2m}\right],
\end{eqnarray}
\begin{eqnarray}
\Psi^{\cal P}(q,Q)&=&\frac{\Psi_0^{\cal P}({\bf q})}
{\left[\frac{\epsilon(q)}{m}\frac{(\epsilon(q)+m)}{2m}\right]}
\left[\frac{\hat v_2-1}{2}+\hat v_2\frac{{\bf q}^2}{2m(\epsilon(q)+
m)}+\frac{\hat{q}}{2m}\right]\cr
&&\times\gamma_5(1+\hat v_2)
\left[\frac{\hat v_2+1}{2}+\hat v_2\frac{{\bf q}^2}{2m(\epsilon(q)+
m)}-\frac{\hat{q}}{2m}\right],
\end{eqnarray}
where $v_1=P/M_{\cal V}$, $v_2=Q/M_{\cal P}$; ${\tilde\epsilon}$ is
the polarization vector of the vector charmonium;
$\epsilon(p)=\sqrt{p^2+m^2}$ and $m$ is the $c$ quark mass. The
vertex function $\Gamma^\nu(p,P;q,Q)$ at leading order in $\alpha_s$
can be written as a sum of four contributions:
\begin{eqnarray}\label{vf}
\Gamma^\nu(p,P;q,Q)&=&
\gamma_\mu\frac{(\hat r-\hat q_1+m)}{(r-q_1)^2-m^2+i\epsilon}
\gamma_\beta D^{\mu\nu}(k_2)+
\gamma_\beta\frac{(\hat p_1-\hat r+m)}{(r-p_1)^2-m^2+i\epsilon}
\gamma_\mu D^{\mu\nu}(k_2)\cr
&&+\gamma_\beta\frac{(\hat q_2-\hat r+m)}{(r-q_2)^2-m^2+i\epsilon}
\gamma_\mu D^{\mu\nu}(k_1)+
\gamma_\mu\frac{(\hat r-\hat p_2+m)}{(r-p_2)^2-m^2+i\epsilon}
\gamma_\beta D^{\mu\nu}(k_1),\qquad
\end{eqnarray}
where the gluon momenta are $k_1=p_1+q_1$, $k_2=p_2+q_2$ and
$r^2=s=(P+Q)^2=(p_-+p_+)^2$, $p_-$, $p_+$ are four momenta of the
electron and positron. The dependence on the relative momenta of
$c$-quarks is present both in the gluon propagator $D_{\mu\nu}(k)$
and quark propagators as well as in the relativistic wave functions.
One of the main technical difficulties in calculating the production
amplitude (\ref{M}) consists in performing angular integrations,
since both gluon and quark propagators in the vertex function
(\ref{vf}) contain angles in the denominators. Therefore we expand
these propagators in the relative momenta. Such expansion leads to
the vertex function containing angles only in numerators and, thus,
the angular integrations can be easily performed.

The inverse denominators of quark propagators expanded in the ratio
of the relative quark momenta $p,q$ to the energy  $\sqrt{s}$ up to
the second order can be expressed as follows:
\begin{equation}
\frac{1}{(r-q_{1,2})^2-m^2}=\frac{1}{Z_{1}}\left[1-\frac{q^2}{Z_{1}}\pm \frac{2(rq)}
{Z_{1}}+\frac{4(rq)^2}{Z_{1}^2}+\cdots\right],
\end{equation}
\begin{equation}
\frac{1}{(r-p_{1,2})^2-m^2}=\frac{1}{Z_{2}}\left[1-\frac{p^2}{Z_{2}}\pm \frac{2(rp)}
{Z_{2}}+\frac{4(rp)^2}{Z_{2}^2}+\cdots\right],
\end{equation}
where the factors $Z_{1}$ and $Z_{2}$ differ only due to the bound state corrections:
\begin{equation}
Z_{1}=\frac{2s+2M_{\cal V}^2-M^2_{\cal P}-4m^2}{4},~~
Z_{2}=\frac{2s+2M_{\cal P}^2-M^2_{\cal V}-4m^2}{4}.
\end{equation}
Corresponding expansions of the gluon propagators in Eq.(5) with the
account of terms of order $O(p^2/s,q^2/s)$ are ($Z=s/4$):
\begin{equation}
\frac{1}{k_{2,1}^2}=\frac{1}{Z}\left[1-\frac{p^2+q^2+2pq}{Z}\pm\frac{(rp)+(rq)}{Z}+
\frac{(rp)^2+(rq)^2+2(rp)(rq)}{Z^2}+\cdots\right].
\end{equation}

We expanded the gluon and quark propagators in the ratio of the
relative quark momenta to the center-of-mass energy $\sqrt{s}$ up to
the second order terms in the production vertex function (5) but
preserved all relativistic factors entering the denominators of the
relativistic wave functions (3), (4). This provides the convergence
of the resulting momentum integrals. Then keeping the terms of
second and fourth order in both variables $p$ and $q$ in the
numerator of Eq.(2) from the relativistic wave functions (3)-(4) and
second order from the expansions of the quark and gluon propagators,
we perform the angular averaging taking into account Eq.(1) and
using the following relation:
\begin{equation}
\int p_\mu p_\nu d\Omega_{\bf p}=-\frac{1}{3}{\bf p}^2
\left(g_{\mu\nu}-\frac{P_\mu P_\nu}{M^2}\right).
\end{equation}
Then we can write
the total production amplitude ${\cal M}$ in the form:
\begin{eqnarray}
{\cal M}(e^+e^-&\to& {\cal P}+{\cal
V})=\frac{256}{9}\pi^2\alpha\alpha_s {\cal Q}_c\frac{\sqrt{4M_{\cal
P} M_{\cal V}}}{s^2u^2(1-u)^2(M_{\cal V}+M_{\cal P})} \bar
v(p_+)\gamma^\beta u(p_-)\epsilon_{\sigma\rho\lambda\beta}v_1^\sigma
v_2^\rho\tilde\epsilon^{\ast~\lambda}\cr && \times\int\frac{d{\bf
p}}{(2\pi)^3}\left(\frac{\epsilon(p)+m}{2\epsilon(p)}\right)\Psi_0^{\cal
V} ({\bf p}) \int\frac{d{\bf
q}}{(2\pi)^3}\left(\frac{\epsilon(q)+m}{2\epsilon(q)}\right)\Psi_0^{\cal
P} ({\bf
q})\left[\frac{T_{13}}{Z_{1}}+\frac{T_{24}}{Z_{2}}\right],\qquad
\end{eqnarray}
where $T_{13}$ originates from the sum of the first and third terms
in the vertex function (5) and $T_{24}$ from the sum of the second
and fourth terms. First, using the Form package \cite{FORM} we
presented $T_{13}$ and $T_{24}$ as a series over the factors
$Z_{1}$, $Z_{2}$, $u=M_{\cal P}/(M_{\cal P}+ M_{\cal V})$,
$\kappa=m/(M_{\cal P}+M_{\cal V})$,
$c(p)=[2m/(\epsilon(p)+m)-1]\equiv -{\bf p}^2/(\epsilon(p)+m)^2$,
$c(q)=[2m/(\epsilon(q)+m)-1]\equiv -{\bf q}^2/(\epsilon(q)+m)^2$.
The resulting expressions are cumbersome so we omit them
here.\footnote{They are available from authors:
apm@physik.hu-berlin.de} Then we performed their simplification by
neglecting the bound state corrections in the denominators $Z_{1}$
and $Z_{2}$ (8). This can be done because the value of $\sqrt{s}$ at
which the experimental data were obtained is essentially larger than
the quark bound state energy. In this approximation which does not
influence the accuracy of the calculation (the corresponding error
in the cross section at the energy $\sqrt{s}=10\div 11$ GeV amounts
0.5 $\%$) we have $Z_{1}\approx Z_{2}\approx s/2$. After such
approximation the total cross section for the exclusive production
of pseudoscalar and vector charmonium states in $e^+e^-$
annihilation is given by the following analytical expression:
\begin{eqnarray}
\sigma(s)&=&\frac{8192\pi^3\alpha^2\alpha_s^2{\cal
Q}_c^2}{2187s^4u^5(1-u)^5} \left\{\left[1-\frac{(M_{{\cal
V}}+M_{{\cal P}})^2}{s}\right] \left[1-\frac{(M_{{\cal V}}-M_{{\cal
P}})^2}{s}\right]\right\}^{3/2}\cr && \times\Biggl[ \int\frac{d{\bf
p}}{(2\pi)^3}\left(\frac{\epsilon(p)+m}{2\epsilon(p)}\right)
\Psi_0^{\cal V}({\bf p})\int\frac{d{\bf
q}}{(2\pi)^3}\left(\frac{\epsilon(q)+m}{2\epsilon(q)}\right)\Psi_0^{\cal
P} ({\bf q})T({\bf p},{\bf q})\Biggr]^2,
\end{eqnarray}
where the function $T({\bf p}, {\bf q})$ can be written as follows:
\begin{eqnarray}
&&\!\!\!\!\!\!\!T({\bf p}, {\bf q})=\sum_{k,l=0}^2 \omega_{kl}c^k(p)c^l(q)+
\frac{(M_{\cal V}+M_{\cal P})^2}{s}\sum_{k,l=0}^2\rho_{kl}c^k(p)c^l(q)\\
&&\quad+\frac{(M_{\cal V}+M_{\cal P})^4}{s^2}\sum_{k,l=0}^2\sigma_{kl}c^k(p)c^l(q)+
\frac{(M_{\cal V}+M_{\cal P})^6}{s^3}\gamma_1c(p)c(q)+
\frac{(M_{\cal V}+M_{\cal P})^8}{s^4}\gamma_2c(p)c(q).\nonumber
\end{eqnarray}
The nonzero values of the coefficients $\omega_{kl}$, $\rho_{kl}$,
$\sigma_{kl}$, $\gamma_{1,2}$ are given explicitly in the
Appendix~A.

\begin{figure}
\centering
\includegraphics{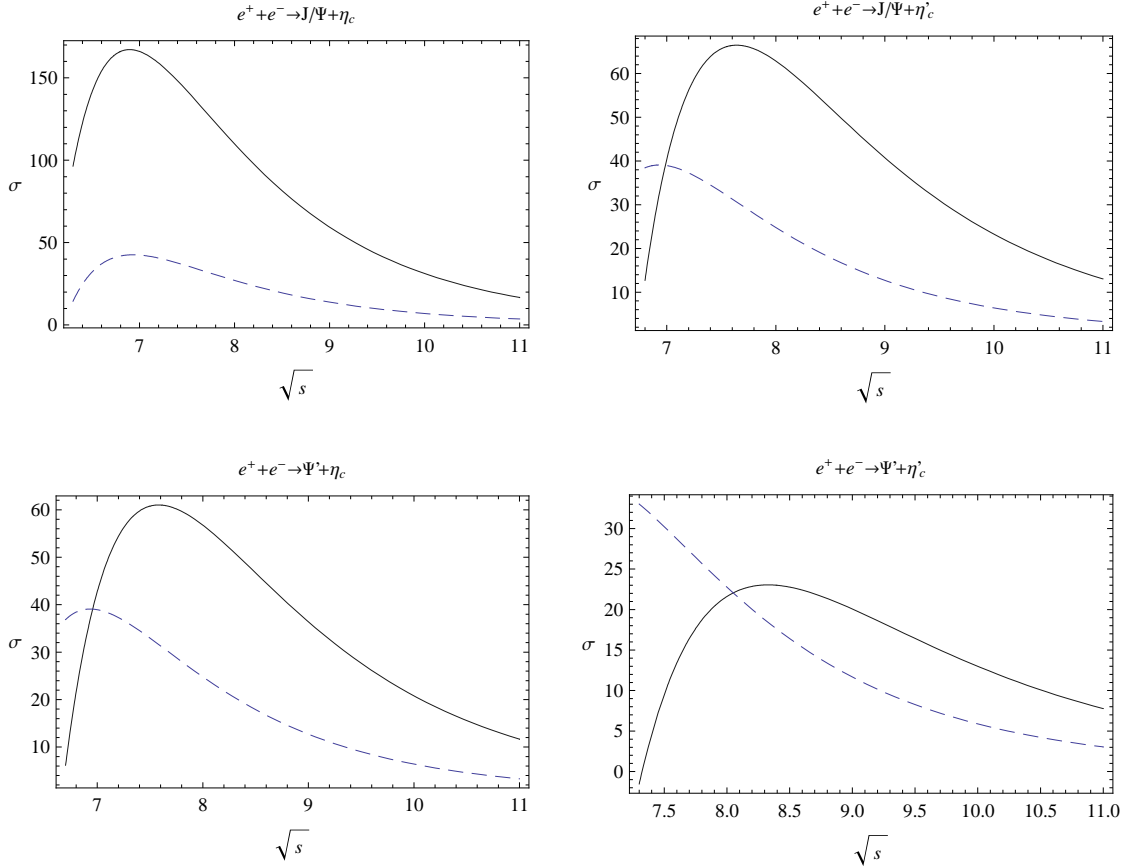}
\caption{The cross section in fb of $e^+e^-$ annihilation into a pair
of $S$-wave charmonium states with the opposite charge parity
as a function of the center-of-mass energy
$\sqrt{s}$ (solid line). The dashed line shows the nonrelativistic result without
bound state and relativistic corrections.}
\end{figure}

The momentum integrals entering Eq.(12) are convergent and we
calculate them numerically, using the wave functions obtained by the
numerical solution of the relativistic quasipotential wave equation
\cite{rqm1,rqm2,rqm3,FFS}. The exact form of the wave functions
$\Psi^{\cal V}({\bf p})$ and $\Psi^{\cal P}({\bf q})$ is extremely
important for getting the reliable numerical results. It is
sufficient to note that the charmonium production cross section
$\sigma(s)$ in the nonrelativistic approximation contains the factor
$|\Psi^{\cal V}_{NR}(0)|^2 |\Psi^{\cal P}_{NR}(0)|^2$. So, small
changes of the numerical values $\Psi^{\cal V}_{NR}(0)$ and
$\Psi^{\cal P}_{NR}(0)$ considerably influence the final result. In
the approach based on nonrelativistic QCD this problem is closely
related to the determination of the color-singlet matrix elements
for the charmonium. Therefore for our calculations we use the
charmonium wave functions $\Psi^{\cal V,P}$ obtained with the
complete nonperturbative treatment of relativistic effects. For this
purpose we consider the quark-antiquark interaction operator
constructed in the relativistic quark model in
Refs.\cite{rqm1,rqm2,rqm3}. Thus, in the present study of the
production amplitude (2) we keep the relativistic corrections of two
types. The first type is determined by several functions depending
on the relative  quark momenta  ${\bf p}$ and ${\bf q}$ arising from
the gluon propagator, the quark propagator and the relativistic
meson wave functions. The second type of corrections originate from
the nonperturbative treatment of the hyperfine interaction in the
quark-antiquark potential which leads to the different wave
functions $\Psi_0^{\cal V}({\bf p})$ and $\Psi_0^{\cal P}({\bf q})$
for the vector and pseudoscalar charmonium states, respectively. In
addition, we systematically accounted the bound state corrections
working with the observed masses of the vector and pseudoscalar
mesons. The calculated masses of vector and pseudoscalar charmonium
states agree well with experimental values \cite{rqm2,PDG}. Note
that all parameters of the model are kept fixed from the previous
calculations of the meson mass spectra and decay widths
\cite{rqm1,rqm2,rqm4}. The masses of the $S$-wave charmonium states
are: $m_{J/\Psi}=3.097$ GeV, $m_{\eta_c}=2.980$ GeV,
$m_{\Psi'}=3.686$ GeV, $m_{\eta_c'}=3.637$ GeV. The strong coupling
constant entering the production amplitude (2) is taken to be
$\alpha_s$=0.21 (see also \cite{BL1,BLL}).

Numerical results and their comparison with several previous
calculations and experimental data are presented in Table I. In
Refs.\cite{He,Bodwin4,Bodwin3} the cross section $\sigma[e^++e^-\to
J/\Psi+\eta_c]$ was calculated with the values 20.04 fb, $17.5\pm
5.7$ fb and $17.6^{+8.1}_{-6.7}$ fb, respectively. The calculated
production cross sections of a pair of $S$-wave charmonium states
are shown in Fig.~2. Our new evaluation of the cross sections in the
reaction $e^++e^-\to {\cal V}_{c\bar c}+ {\cal P}_{c\bar c}$
evidently shows that the systematic account of all relativistic
effects connected with the bound state wave functions, the gluon and
quark propagators removes the discrepancy between theory and
experiment. Numerically, the increase of the cross section $\sigma$
(12) is determined approximately by the factor of 2 coming from the
relativistic corrections entering in the production amplitude (2)
(in this part our results agree with the previous calculations in
Ref.\cite{EM2006}) and by another factor of 2 from the relativistic
bound state wave functions. In our analysis we use the exact
expressions (3)-(4) for the relativistic wave functions. Thus we
correctly take into account all relativistic contributions of orders
$O(v^2)$ and $O(v^4)$ since they are determined by the convergent
momentum integrals due to the presence of the relativistic factors
in the denominators of expressions (3)-(4). Therefore the resulting
theoretical uncertainty is connected with the omitted terms of the
employed truncated expansions (6), (7), (9) which are of order $v^2
p^2/s$. Taking into account that the average value of the heavy
quark velocity squared in the charmonium is $<v^2>=0.3$, we expect
that they should not exceed 5-10\% in the interval of energies
$\sqrt{s}=7\div 11$ GeV. We should remind also that our relativistic
quark model has the phenomenological structure and differs
significantly from the approach of nonrelativistic QCD (NRQCD).
Despite the fact that it is based on the quantum field-theoretic
approach, it contains a number of the phenomenological parameters
which we fixed solving many tasks in the quarkonium physics.
Unfortunately, we can not control the theoretical accuracy in the
same manner as in NRQCD. We obtained the theoretical predictions for
the masses and decay rates of different charmonium states with more
than one per cent accuracy. So, we suppose in this study that there
are no additional essential theoretical uncertainties in the bound
state wave functions connected with the formulation of our model in
regions of nonrelativistic and relativistic momenta.

\begin{table}
\caption{\label{t1} Comparison of the obtained results with previous theoretical
predictions and experimental data.}
\bigskip
\begin{ruledtabular}
\begin{tabular}{|c|c|c|c|c|c|c|c|c|}   
State  & $\sigma_{BaBar}\times$ & $\sigma_{Belle}\times $ &
$\sigma$ $(fb)$ &$\sigma_{NRQCD}$& $\sigma$ $(fb)$ &$\sigma$ $(fb)$ & $\sigma$ $(fb)$ &
Our result \\
$H_1H_2$   &$ Br_{H_2\to charged\ge 2}$ & $Br_{H_2\to charged\ge 2}$ &
\cite{BLL} &$(fb)$ \cite{BL1}& \cite{Chao1}  &\cite{BL1}& \cite{EM2006}  &  (fb)  \\
   &  $(fb)$ \cite{BaBar} & $(fb)$ \cite{Belle} &   &  &  &    &    &  \\   \hline
$\Psi(1S)\eta_c(1S)$ & $17.6\pm 2.8^{+1.5}_{-2.1}$ & $25.6\pm 2.8\pm
3.4$ & 26.7 &  3.78& 5.5 & 7.4  & 7.8  &  $22.2\pm 3.6$ \\  \hline
$\Psi(2S)\eta_c(1S)$ &  & $16.3\pm 4.6\pm 3.9$ & 16.3 &  1.57& 3.7 &
6.1  & 6.7  &  $15.3\pm 2.4$ \\  \hline $\Psi(1S)\eta_c(2S)$ &
$16.4\pm 3.7^{+2.4}_{-3.0}$ & $16.5\pm 3.\pm 2.4$ & 26.6 &  1.57
&3.7 &7.6 & 7.0  &  $16.4\pm 2.6$ \\  \hline $\Psi(2S)\eta_c(2S)$ &
& $16.0\pm 5.1\pm 3.8$ &
14.5 &  0.65& 2.5 &5.3 &5.4  &  $9.6\pm 1.5$ \\  
\end{tabular}
\end{ruledtabular}
\end{table}
It is important to point out that it is not possible to simply
compile the enhancements of the production cross sections
originating from our calculation of the relativistic contributions
and from the one-loop corrections calculated in Ref.\cite{ZGC}. The
latter was done in the nonrelativistic limit. Indeed in our model
the interaction potential in the relativistic wave equation contains
the one-loop radiative corrections. Therefore the inclusion of the
one-loop corrections considered in \cite{ZGC} in our calculation
requires their complete recalculation using our relativistic wave
functions, since we take into account effectively some part of the
one-loop diagrams connected with the exchange of gluons between
heavy quarks in the final state \footnote{This is beyond the scope
of the present paper.}. As a result both nonperturbative and
partially perturbative contributions are taken into account.

Thus our approach cannot be directly confronted with the one of
Ref.\cite{ZGC}. The radiative corrections are the main source of the
theoretical uncertainty in our calculations. Indeed, available
estimates of one-loop corrections in the nonrelativistic limit
indicate that they are considerable. Taking their values from
\cite{ZGC} (relativistic factor $K=1.8$ to nonrelativistic result)
we estimate that this part of the theoretical error should not
exceed 15\%. Therefore the total theoretical uncertainty amounts to
16\% for the energy region $\sqrt{s}={10.6}$ GeV. To obtain this
estimate we add the above mentioned relativistic and one-loop
uncertainties in quadrature (as it was done in Ref.\cite{Bodwin4}).
These theoretical errors in the calculated production cross section
at $\sqrt{s}=10.6$ GeV are shown directly in Table I. There are no
additional uncertainties related to the choice of $m_c$ or any other
parameters of the model, since their values were fixed from our
previous consideration of meson and baryon properties
\cite{rqm1,rqm2,rqm3,rqm4,rqm5}.

In summary, we presented a systematic treatment of relativistic
effects in the double charmonium production in $e^+e^-$
annihilation. We explicitly separated two different types of
relativistic contributions to the production amplitudes. The first
type includes the relativistic $v/c$ corrections to the wave
functions and their relativistic transformations which were for the
first time exactly taken into account. The second type includes the
relativistic $p/\sqrt{s}$ corrections emerging from the expansion of
the quark and gluon propagators. The latter corrections were taken
into account up to the second order. It is important to note that
the expansion parameter $p/\sqrt{s}$ is very small. Contrary to the
previous calculations within NRQCD all obtained expressions for the
relativistic contributions are now expressed through converging
integrals. Thus no additional uncertainty related to their
regularization emerges. Therefore we can reliably estimate the
uncertainty originating from the neglected higher-order relativistic
contributions. The calculated values for the production cross
sections agree well with experimental data.

\acknowledgments The authors are grateful to participants of the
ITEP Conference ``Physics of Fundamental Interactions'' (section
``Heavy quark physics'') for general discussions and fruitful
remarks. One of the authors (A.P.M.) thanks M.M\"uller-Preussker and
the colleagues from the Institute of Physics of the Humboldt
University in Berlin for warm hospitality. The work is performed
under the financial support of the {\it German Academic Exchange
Service} (DAAD) (A.P.M.), Russian Science Support Foundation
(V.O.G.), Russian Foundation for Basic Research (RFBR) (grant
No.08-02-00582) (R.N.F. and V.O.G.).

\appendix

\section{The coefficients $\omega_{ij}$, $\sigma_{ij}$, $\rho_{ij}$, $\gamma_i$
entering in the production cross section}
\begin{equation}
\omega_{00}=-18 (-1+2\kappa-3 u) (u-1)^2 u^2,
\end{equation}
\begin{equation}
\omega_{01}=6 (u-1)^2 \left[32 {\kappa}^3+16 {\kappa}^2 \left(-5+u\right)-6
{\kappa} u^2+3 \left(1-5 u\right) u^2\right],
\end{equation}
\begin{equation}
\omega_{10}=6 u^2 \left[96 {\kappa}^3-34 {\kappa} \left(u-1\right)^2+\left(u-1\right)^2
\left(5 u-1\right)-16 {\kappa}^2 \left(1+11 u\right)\right],
\end{equation}
\begin{equation}
\omega_{11}=-2 \Bigl[536 {\kappa}^5+102 {\kappa} \left(u-1\right)^2 u^2+3 \left(u-1\right)^2 u^2
\left(1+3 u\right)-8 {\kappa}^4 \left(61+67 u\right)-
\end{equation}
\begin{displaymath}
-6 {\kappa}^3 \left(114+u \left(161 u-228\right)\right)-2 {\kappa}^2 \left(46+u \left(-170+u
\left(175+237 u\right)\right)\right)\Bigr],
\end{displaymath}
\begin{equation}
\omega_{12}=-96\kappa
u^2\left[6\kappa^2-2(u-1)^2-\kappa(1+11u)\right],
\end{equation}
\begin{equation}
\omega_{21}=-96\kappa^2 (u-1)^2\left(-5+2\kappa+u\right),
\end{equation}
\begin{equation}
\gamma_1=64 {\kappa}^3 \Bigl[64 {\kappa}^6 \left(5+2 u \left(2 u-5\right)\right)+32 {\kappa}^5
\left(4+\left(u-1\right) u \left(14+3 u\right)\right)+
\end{equation}
\begin{displaymath}
+4 {\kappa}^2 (u-1)^2 \left(146+u \left(-584+u \left(851+6 u \left(21 u-89\right)\right)\right)\right)
-16 {\kappa}^4\times
\end{displaymath}
\begin{displaymath}
\left(44+u \left(u\left(273+u \left(49 u-194\right)\right)-176\right)\right)
-16 {\kappa}^3 \left(7+u \left(u \left(37+u \left(u-25+15 u^2\right)\right)-27\right)\right)-
\end{displaymath}
\begin{displaymath}
-(u-1)^2 \left(194+u \left(-1164+u \left(2754+u \left(-3256+3 u \left(676+27 \left(u-8\right) u\right)
\right)\right)\right)\right)+
\end{displaymath}
\begin{displaymath}
+2 {\kappa} (u-1) \left(2+u \left(-108+u \left(523+u \left(-992+u \left(874+3 u \left(9 u-106\right)
\right)\right)\right)\right)\right)\Bigr],
\end{displaymath}
\begin{equation}
\gamma_2=-128 {\kappa}^4 (1-2 u)^2 \Bigl[-31+32 {\kappa}^5+2 {\kappa} (1-3 u)^2 (1+u)^2-16 {\kappa}^4
(1+3 u)-
\end{equation}
\begin{displaymath}
-16 {\kappa}^3 \left(1+u \left(5 u-2\right)\right)+8 {\kappa}^2 \left(7+u \left(-17+u \left(11+15
u\right)\right)\right)+
\end{displaymath}
\begin{displaymath}
+u \left(151-u \left(286+3 u \left(-98+u \left(55+9 u\right)\right)\right)\right)\Bigr],
\end{displaymath}
\begin{equation}
\sigma_{01}=96 {\kappa}^2 (u-1)^2 \Bigl[-17+2 {\kappa} (1-2 u)^2+16 {\kappa}^4 (-1+u)-
\end{equation}
\begin{displaymath}
-8 {\kappa}^2 (u-1) (4+u (5 u-8))+u \left(81+u \left(-144+u \left(124-57 u+9 u^2\right)\right)\right)\Bigr],
\end{displaymath}
\begin{equation}
\sigma_{10}=48 {\kappa}^2 u^2 \Bigl[-1+32 {\kappa}^5+16 {\kappa}^4 (1-5 u)-16 {\kappa}^3 (1+u (5 u-2))+
\end{equation}
\begin{displaymath}
+8 {\kappa}^2 (5 u-1) (1+u (5 u-2))+6 {\kappa} \left(1+u \left(3 u-2\right) \left(2+u \left(2+u\right)\right)
\right)-
\end{displaymath}
\begin{displaymath}
-u \left(7+u \left(-34+u \left(2+51 u+45 u^2\right)\right)\right)\Bigr],
\end{displaymath}
\begin{equation}
\sigma_{11}={\kappa}^2 \Bigl[600-8 {\kappa} \left(-1695+2 {\kappa} \left(-45+{\kappa} \left(1527+2 {\kappa}
\left(11+{\kappa} \left(-659+2 {\kappa} \left(67 {\kappa}-31\right)\right)\right)\right)\right)\right)-
\end{equation}
\begin{displaymath}
-4328 u+16 {\kappa} \left(-5085+{\kappa} \left(-769+2 {\kappa} \left(3054+{\kappa} \left(-351+2 {\kappa}
\left(97 {\kappa}-659\right)\right)\right)\right)\right) u+
\end{displaymath}
\begin{displaymath}
+4 \left(3235+{\kappa} \left(49319+4 {\kappa} \left(2783+{\kappa} \left(-9443+{\kappa} \left(1071+1681
{\kappa}\right)\right)\right)\right)\right) u^2-
\end{displaymath}
\begin{displaymath}
-4 (4987+4 {\kappa} (15419+{\kappa} (3801
+667{\kappa} ({\kappa}-10)))) u^3+2 \Bigl(7933+{\kappa} \Bigl(83683+
\end{displaymath}
\begin{displaymath}
+2 \left(7829-8265 {\kappa}\right) {\kappa}\Bigr)\Bigr) u^4-2 \left(3711+29142 {\kappa}
+3770 {\kappa}^2\right) u^5+(4765+9851 {\kappa}) u^6+983 u^7\Bigr],
\end{displaymath}
\begin{equation}
\rho_{01}=96 {\kappa}^2 (u-1)^2 \left[16+12 {\kappa}^2 (u-1)+u (-40+(33-13 u) u)-2 {\kappa} (2+(u-4) u)\right],
\end{equation}
\begin{equation}
\rho_{10}=24 {\kappa} u^2 \Bigl[48 {\kappa}^4+16 {\kappa}^3 (1-7 u)+{\kappa}^2 \left(-4+8 u-76 u^2\right)-
\end{equation}
\begin{displaymath}
-(u-1)^2 (-7+u (14+3 u))+2 {\kappa} \left(-1+u \left(-1+u+73 u^2\right)\right)\Bigr],
\end{displaymath}
\begin{equation}
\rho_{11}=-8 {\kappa} \Bigl[528 {\kappa}^6-128 {\kappa}^5 (3+5 u)+3 (u-1)^2 u^2 (-7+u (14+3 u))-12
{\kappa}^4 (67+2 u (56 u-67))+
\end{equation}
\begin{displaymath}
+4 {\kappa}^3 \left(97+u \left(-259+6u(39+9u)\right)\right)+{\kappa}^2 (814+u (-3256+u (5017+u
(-3522+1159 u))))+
\end{displaymath}
\begin{displaymath}
+{\kappa} \left(70+u \left(-382+u \left(969+u \left(-1323+2 u(469+80)\right)\right)\right)\right)\Bigr],
\end{displaymath}
$\rho_{21}=-\rho_{01}$, $\rho_{12}=-\rho_{10}$,
$\sigma_{21}=-\sigma_{01}$, $\sigma_{12}=-\sigma_{10}$.

\end{document}